\documentclass[10pt]{iopart}

\usepackage{iopams}
\usepackage{graphicx}
\usepackage{bm}
\begin{document}

\title{Height distribution of equipotential lines in a region confined by a rough conducting boundary}

\author{C. P. de Castro}
\address{Instituto de F\'{\i}sica, Universidade Federal da Bahia,
   Campus Universit\'{a}rio da Federa\c c\~ao,
   Rua Bar\~{a}o de Jeremoabo s/n,
40170-115, Salvador, BA, Brazil}
\ead{caioporto@ufba.br}

\author{T. A. de Assis}
\address{Instituto de F\'{\i}sica, Universidade Federal da Bahia,
   Campus Universit\'{a}rio da Federa\c c\~ao,
   Rua Bar\~{a}o de Jeremoabo s/n,
40170-115, Salvador, BA, Brazil}
\ead{thiagoaa@ufba.br}

\author{C. M. C. de Castilho}
\address{Instituto de F\'{\i}sica, Universidade Federal da Bahia,
   Campus Universit\'{a}rio da Federa\c c\~ao,
   Rua Bar\~{a}o de Jeremoabo s/n,
40170-115, Salvador, BA, Brazil}
\address{Instituto Nacional de Ci\^{e}ncia e Tecnologia em Energia e
Ambiente, Universidade Federal da Bahia
   Campus Universit\'{a}rio da Federa\c c\~ao,
40170-115, Salvador, BA, Brazil}
\ead{caio@ufba.br}

\author{R. F. S. Andrade}
\address{Instituto de F\'{\i}sica, Universidade Federal da Bahia,
   Campus Universit\'{a}rio da Federa\c c\~ao,
   Rua Bar\~{a}o de Jeremoabo s/n,
40170-115, Salvador, BA, Brazil}
\ead{randrade@ufba.br}

\begin{abstract}

This work considers the behavior of the height distributions of the
equipotential lines in a region confined by two interfaces: a
cathode with an irregular interface and a distant flat anode. Both
boundaries, which are maintained at distinct and constant potential
values, are assumed to be conductors. The morphology of the cathode
interface results from the deposit of $2 \times 10^{4}$ monolayers
that are produced using a single competitive growth model based on
the rules of the Restricted Solid on Solid and Ballistic Deposition
models, both of which belong to the Kadar-Parisi-Zhang (KPZ)
universality class. At each time step, these rules are selected with
probability $p$ and $q = 1 - p$. For several irregular profiles that
depend on $p$, a family of equipotential lines is evaluated. The
lines are characterized by the skewness and kurtosis of the height
distribution. The results indicate that the skewness of the
equipotential line increases when they approach the flat anode, and
this increase has a non-trivial convergence to a delta distribution
that characterizes the equipotential line in a uniform electric
field. The morphology of the equipotential lines is discussed; the
discussion emphasizes their features for different ranges of $p$
that correspond to positive, null and negative values of the
coefficient of the non-linear term in the KPZ equation.
\end{abstract}

\pacs{68.55.-a, 68.35.Ct, 81.15.Aa , 05.40.-a}

\maketitle

\section{Introduction}
\label{intro}

Growth phenomena in non-equilibrium conditions~\cite{Barabasi,Krug}
are an important subject of condensed matter physics. Many
properties of specific devices depend on rough surfaces that are
produced under such non-equilibrium conditions. Technological
applications such as conducting-field emitter devices that operate
in Ultra High Vacuum (UHV) conditions, require the control of
several properties such as the surface morphology. It is well known
that large values of the current intensity on field emitter devices
are connected to extremely small values of the effective emitting
area \cite{Forbes}. One method to increase the effective area of
emission is to use surfaces with an ending fractal boundary with
self-affine scale invariance instead of considering structurally
regular tips. In fact, the connection between the morphology of
irregular surfaces including sharp conducting tips and the field emission properties has been a
subject of intense research \cite{Du,deassis}. In a recent work \cite{Cabrera}, Cabrera \textit{et al.} measured
experimentally the current-voltage (I-V) of a tunnel junction consisting
of a sharp electron emitting metallic tip at a variable distance (``$\Delta$") from a planar collector and emitting electrons using electric field assisted
emission. Their results showed scale invariance of
the tunnel junction with respect to changes in $\Delta$
(from few $mm$ to several $nm$), which means that
the physical laws governing the flow of current are invariant with
respect the changes of the length scale $\Delta$.
However, this behavior fails when $\Delta$ was only a few nanometers
as showed in reference \cite{Kyritsakis}.
The reported scale invariance for a single
tip can be regarded as a preliminary study of the more general problem of field emission by
an irregular conducting interface that will be considered in this work.

In the statistical mechanics framework, macroscopic laws that emerge
in surface growth can be explained at different spatial and temporal
scales considering a theoretical microscopic description that
involves simple probabilistic laws. An interesting related point is
the possibility of modeling these systems so that in the large-scale
limit (i.e., where the scale invariance arises), they do not depend
on the corresponding details but only on symmetries and the
corresponding conservation laws.

Previous theoretical works related to the emitting properties of
rough surfaces \cite{Cajueiro, Hugo, Assis1, Assis2, Budaev} have
studied the behavior of the electric potential in a region bounded
by a rough conducting profile/surface and a smooth conducting
line/plane, which is held to a constant voltage bias. In particular,
the references \cite{Hugo, Assis1} analyzed the roughness exponent
of the equipotential lines (surfaces) in 2 (3)-dimensional systems.
However, to our knowledge, no previous work discussed the properties
of equipotential lines/surfaces under the perspective of its height
distribution properties.

In this work, we investigate the height distribution properties of a
family of equipotential lines in a region that is confined by a
rough profile, which is regarded as an electron-emitting cathode,
and a flat anode which is placed sufficiently away from it. The
profile results from a numerical simulation of the deposition and
growth surface models, whereas the cathode and the anode are subject
to an electric potential difference. Our results are mainly based on
the model that was introduced by Silva and Moreira
\cite{Silva1,Silva2}, which is compared to those obtained for the
simpler Family model \cite{family} in some limiting cases. We
emphasize two consequences of our results for the following aspects:
the height distribution presents a more detailed characterization of
the equipotential lines than that provided by the roughness
exponents and such investigation of equipotential lines offers a
much more reliable method to characterize the surface. For
particular applications, our work can contribute to elucidate the
image distortions in probe microscopies such as Electrostatic Force
Microscopy (EFM), where the knowledge of the electric field
distribution along the tip probe is relevant to extract
morphological information about the real irregular surface. This
main conclusion from our work is particularly useful to study
surfaces under the influence of non-linear mechanisms in the growth
phase, where the surface width is not expressed as a power law with
a single growth exponent for a large number of deposited monolayers.

This investigation is also motivated by the quoted experimental
aspects related to the electronic properties of rough surfaces and
the recent theoretical advances regarding the solution of the
Kardar-Parisi-Zhang (KPZ) equation \cite{KPZ} in d = 1+1,
where, in general, $d=n+1$ indicates that
a system composed by a n-dimensional substrate (here, 1-dimensional)
and an extra dimension to where an important effect produced by the
substrate is propagated.

The remainder of the paper is organized as follows. In Sec.
\ref{SEC0}, we describe the main features of the used models and
emphasize the conditions under which they reproduce the typical
features of the KPZ and Edwards-Wilkinson (EW) \cite{EW} growth
dynamics. In Sec. \ref{SECI}, we describe the methodology to
calculate and characterize of the equipotential lines. In Sec.
\ref{SECII}, we discuss our results for the morphological properties
of equipotential lines and analyze the corresponding behavior of the
skewness and kurtosis with the electric potential. In Sec.
\ref{conclusion}, we summarize the results and present our
conclusions.

\section{Models}
\label{SEC0}

It is convenient to discuss the properties of the computational
surface growth models that are used in this work with those of the
KPZ equation in $1+1$ dimensions

\begin{equation}
\partial_{t}h(x,t)= \mu_{0} + \nu \triangle h + \frac{\lambda}{2} (\nabla h)^2 + \eta(x,t),
\label{Eq00}
\end{equation}
where $h(x,t)$ represents the height of the profile in the $y$ direction with
respect to the line, $h(x,t=0)=0$, that describes the flat profile at $t=0$.
Eq. \ref{Eq00} comprises several terms that consider the essential
features of a stochastic surface growth. These features include
irreversibility and locality, which enhance the surface roughness.
In eq. \ref{Eq00} $\mu_{0}$ corresponds to a constant driving force,
$\nu$ is the diffusion coefficient of the deposited atoms on the
surface, and $\eta(x,t)$ is a white Gaussian noise that mimics the
stochastic nature of the growth process. The non-linear term
accounts for a considerable number of experimental results
\cite{Takeuchi,Yunker} for the height distribution. Many of them
cannot be described by a Gaussian function, which is the solution of
the linear EW equation, which corresponds to setting $\lambda=0$ in
eq. \ref{Eq00}.

The investigated model depends on a parameter $p$, which describes
the probability that the deposited particle follows the Restricted
Solid on Solid (RSOS) local rule, whereas $q=1-p$ is the probability
related to the Ballistic Deposition (BD) rule. Both RSOS and BD
discrete models belong to the universality class that is defined by
the KPZ equation but are characterized by negative and positive
values of the parameter $\lambda$, respectively. Although the used
model has not been proved to be equivalent to a KPZ equation, recent
numerical investigations suggest that a change in the probability
$p$ changes the value of $\lambda$ in a KPZ equation, which may be
suitable to describe the competitive model \cite{Muraca,Reis}.

For almost all ranges of $p$, a growing profile first experiences a
transient phase before attaining a stationary growth regime, where
its width $W$ scales with the number of monolayers with the typical
exponent $\beta_{KPZ}=1/3$ of the KPZ universality class. For finite
profiles, this regime is interrupted when $W$ exponentially relaxes
to a saturation regime and becomes constant. In this work, we
analyze only rough interfaces in the growth regime, which
corresponds to typical experimental conditions. In the first
transient phase, depending on $p$, the growth exponent can reach
values close to those typical of EW dynamics. Starting from $p=0$,
the length of the transient phase increases until a specific value
$p^*\approx0.83$, where the EW typical value is always valid. For
$p>p^*$, the model reaches KPZ scaling again after a decreasing
transient time \cite{Oliveira2}. Because the condition $p=p^*$ is
very special, we find it wise to compare the properties of the
profiles and equipotential lines with those obtained when the
cathode is represented by Family-model produced profile
\cite{family}, which shows the EW properties notably clearly.

The interesting behavior of the competitive model makes it suitable
to investigate how the equipotential lines change when $p$ is
selected to reproduce either KPZ or EW dynamics, and for a fixed
value of $p$, how the magnitude of the non-linear terms affects the
EW-KPZ crossover as a function of time \cite{Oliveira2}, which is
particularly important for experimental interests. Indeed, the
competition between physical mechanisms is present in many real
processes in thin-film science. Then, the actual study clearly
advances on the previously one where single models were considered,
and the roughness exponent of equipotential lines were evaluated
without a perspective for height distribution properties.

The distribution of height fluctuations of the KPZ equation, for a
flat initial condition, can be expressed by Gaussian Orthogonal
Ensemble (GOE) Tracy-Widom (TW) distributions~\cite{Spohn}.
Therefore our results can be better appreciated by comparing the
skewness ($S$) and kurtosis ($K$) of the equipotential height
distributions with the typical TW values for the corresponding
quantities. Both $S$ and $K$ depend on the average distance from the
equipotential lines to the cathode and anode. Such dependence has
intrinsic features according to whether $p<p^{*}$ or $p>p^{*}$. A
third distinct behavior is observed at $p=p^{*}$, when the profile
falls into an EW universality class.

We emphasize again that our investigation opens the possibility of
analyzing the effects on the behavior of equipotential lines of
distinct surfaces with growth exponents near those that feature an
EW class. This possibility is important because of the emergence of
local growth exponents in the first transient growth phase, which is
near the phase that characterizes the EW dynamics when $p
\thickapprox p^{*}$. As shown in Sec. \ref{SECII}, our results
indicate that the height distributions of equipotential lines for
such transient patterns are clearly different from those obtained
for an actual EW profile , no matter whether they are produced using the
competitive model at $p=p^{*}$ or the Family model.

\section{Methodology: Determination and Characterization of Equipotential Lines}
\label{SECI}

The current investigation starts by constructing the rough profile
$h(x)$, which acts as the cathode and results from one of the quoted
deposition models. Here, $x$ and $y$ are integer numbers
that represent distance measured in terms of some basic unit distance $u$, the precise size
of which is not important in a theory of the kind under discussion.
The growth process along the $y$ direction of the considered $1 + 1$ system
(one-dimensional substrate + height, as mentioned in Sec.\ref{intro})
starts with a flat substrate $h(x,t=0)=0$, where $x$ is an integer
number, and $x\in [1,L]$. The column height $h(x)$ is the $y$
coordinate of the topmost adatom at position $x$. We have used
profiles with lateral size up to $L=10^{6}$ and up to $T = 2 \times
10^{4}$ deposited monolayers (ML). For such parameter values,
finite-size effects no longer influence the interpretation of our
results and can be neglected. During the growth process, the number
of monolayers is also referred to as the integration time $t$; in
this way, each time unit corresponds to the deposition of $L$
particles. For the convenience of using a unified notation for all
profiles and equipotential lines, let us switch from $h(x)$ to
$h(x,t)=h_{p,\phi}(x,t)$, where we emphasize the dependence of
$h(x)$ on $t$. The subscript $p$ stands for the probability of
choosing the RSOS rule in the competitive model. Finally, the
subscript $\phi$ is used to identify the constant value of the
equipotential line. In this work, $\phi=0$ and $\phi=A$ correspond
to the cathode and anode surfaces, respectively.

The flat anode, which is defined by $h_{p,A}$, is
placed at a distance $\langle d \rangle$ away from the cathode. $\langle d \rangle$ is measured by
the number of vertical spacings between the average height of the
profile $\overline{h}_{p,0}=\frac{1}{L} \sum\limits_{x=1}^{L}
\left[h_{p,0}(x,T)\right]$ and the flat anode, i.e:

\begin{equation}
\langle d \rangle = h_{p,A} - \overline{h}_{p,0}.
\label{EqE}
\end{equation}
As previously indicated,
$\overline{h}_{p,0}$ is evaluated at $t=T$, except if explicitly
indicated.

To compare the results for different rough profiles,
we fixed the value of $\langle d \rangle$. As a consequence $h_{p,A}$, which
is measured with respect to the substrate where the film is grown, changes
with $p$.
In this way, to select a proper and fixed value of $\langle d \rangle$, such that for any value of $p$ the
anode is sufficiently away from the cathode, we first determined the
value of $p$ for which the roughness $W$ (defined in eq. \ref{Eq2a}) attains its maximum, which is $p=0$ when $W\approx 37.3$. Next, we identified the vertical coordinate
of the highest peak $h^{p=0}_{max}$ for this value of $p$ and set
$h_{0,A} = h^{p=0}_{max}+\Delta$, which result, using eq. \ref{EqE}, in:

\begin{equation}
\langle d \rangle = h^{p=0}_{max} + \Delta - \overline{h}_{0,0},
\label{EqE1}
\end{equation}
where $\Delta$ corresponds to the distance
defined on the second paragraph of Sec.\ref{intro}.
At this point, let us comment on the reasons to choose the particular
 value of $\Delta$ used in this work.
For $\Delta >> h^{p=0}_{max}$, the results are expected
to be largely independent of the precise value of $\Delta$,
but the computational work is largely increased. We have observed that,
for all the rough profiles, the equipotential lines become flat and parallel to
the $x$ axis very rapidly after the value of $y$ surpasses the highest peak. Therefore,
we choose the value $\Delta=10$, which corresponds to $\langle d \rangle=198$. We use this average distance for all $p$, because it is already sufficient large to have
very smooth equipotential lines (with roughness exponent, $\alpha$, near to unity \cite{Assis1}) that are observed at the flat anode.
If we consider the basic unit distance $u=5$ nm, we obtain $\Delta=50$ nm for $p=0$, and $\langle d \rangle \simeq 1$ $\mu$m,
which correspond to the typical scale used in the non-contact mode in atomic force microscopy (like EFM) where the electrostatic force is probed.
Then, the results of this work can motivate experimental realizations in this regime.

Observed at this angle, the geometrical set up depends on two
different scales (expressed by $L$ and $H$) along the horizontal and
vertical directions. However, the rough profile has its own length
scale in the vertical direction, which is the roughness

\begin{equation}
W_{p,0}(L,t)= \left[L^{-1} \sum\limits_{x=1}^{L} \left(h_{p,0}(x,t) - \overline{h}_{p,0}(t) \right)^{2}\right]^{1/2}
\label{Eq2a}
\end{equation}
\noindent where $\overline{h}_{p,0}(t) = L^{-1}
\sum\limits_{x=1}^{L} [h_{p,0}(x,t)]$.

When $t=T$, the profiles that are grown using the competitive model
are in the asymptotic growth regime ($W\sim t^{\beta_{KPZ}}$) for $p
\in [0,0.75) \bigcup (0.9,1]$. This regime corresponds to typical
experimental conditions, in contrast to the exponential relaxation
that is observed in steady-state (long-time) properties. For
$p \in [0.75, p^{*})\bigcup(p^{*},0.90]$, the asymptotic growth KPZ
regime has not yet been reached, and the system is in the EW-KPZ
crossover \cite{Caio,Oliveira}. For $p=p^{*}\approx 0.83$, the
system is in the critical point and the corresponding growth regime
is such that $W\sim t^{\beta_{EW}}$, where $\beta_{EW} =1/4$ is the
EW growth exponent.

The next step is to compute the equipotential lines,
$h_{p,\phi}(x)$. Hence, we numerically solve Laplace's equation

\begin{equation}
\nabla^{2} \phi = 0,
\label{Eq2}
\end{equation}
for the electric potential $\phi$ in the region between the two
conductors, which are held at constant potential values of
$\phi^{C}$ and $\phi^{A}$ for the rough cathode and the flat anode
that is parallel to the $x$ axis, respectively. The equation is
solved using Liebmann's method and the appropriate Dirichlet
conditions. The geometrical characterization of the equipotential
lines is insensitive to the choice of $\phi^{C}$ and
$\phi^{A}\neq\phi^{C}=0$. However, to directly compare some
experimental devices, we choose the typical values $\phi^{C}=0$ and
$\phi^{A}=10^{3}$.

\begin{figure}[!htb]
\begin{center}
\includegraphics [width=8.5cm] {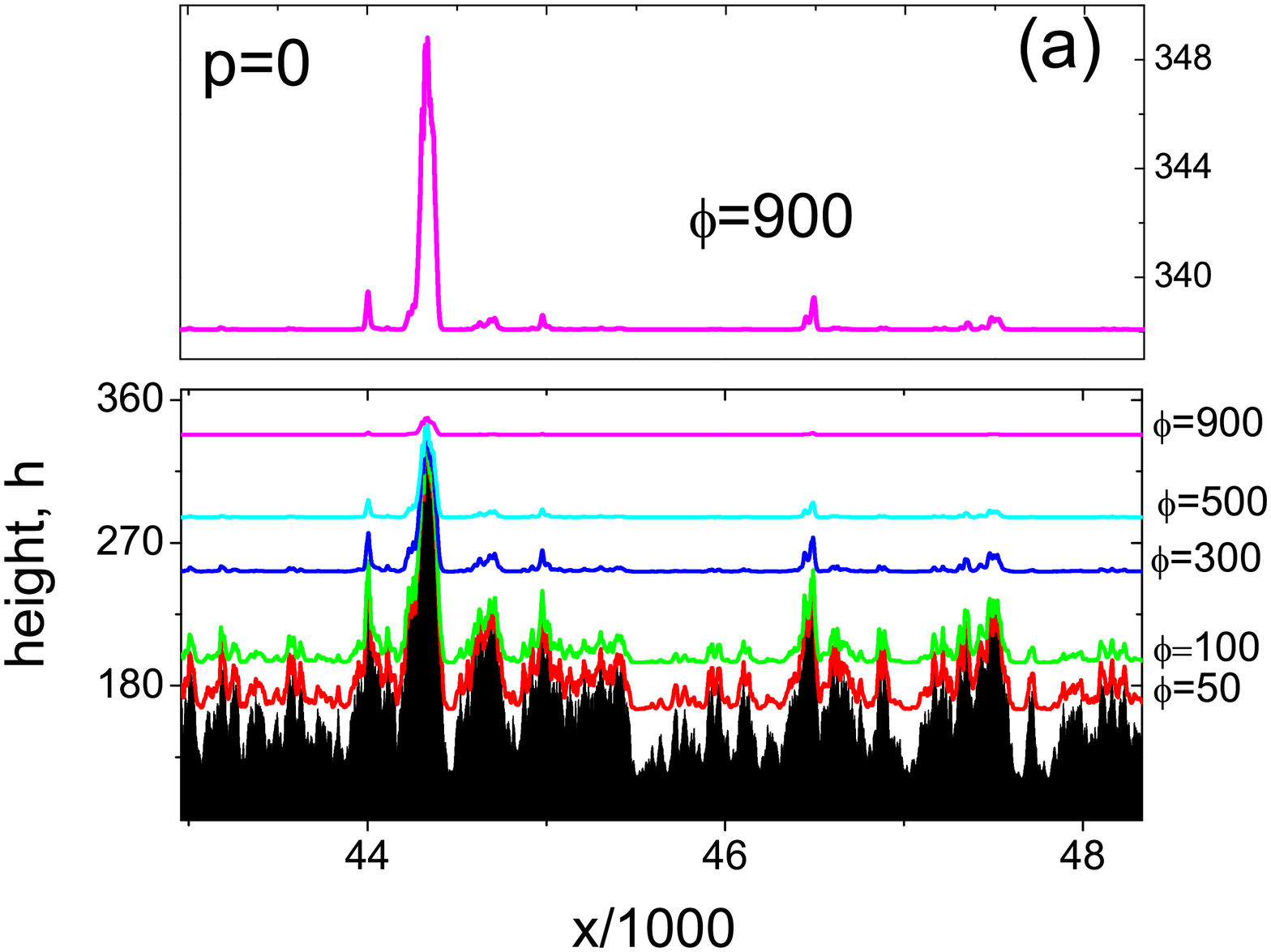}
\includegraphics [width=8.5cm] {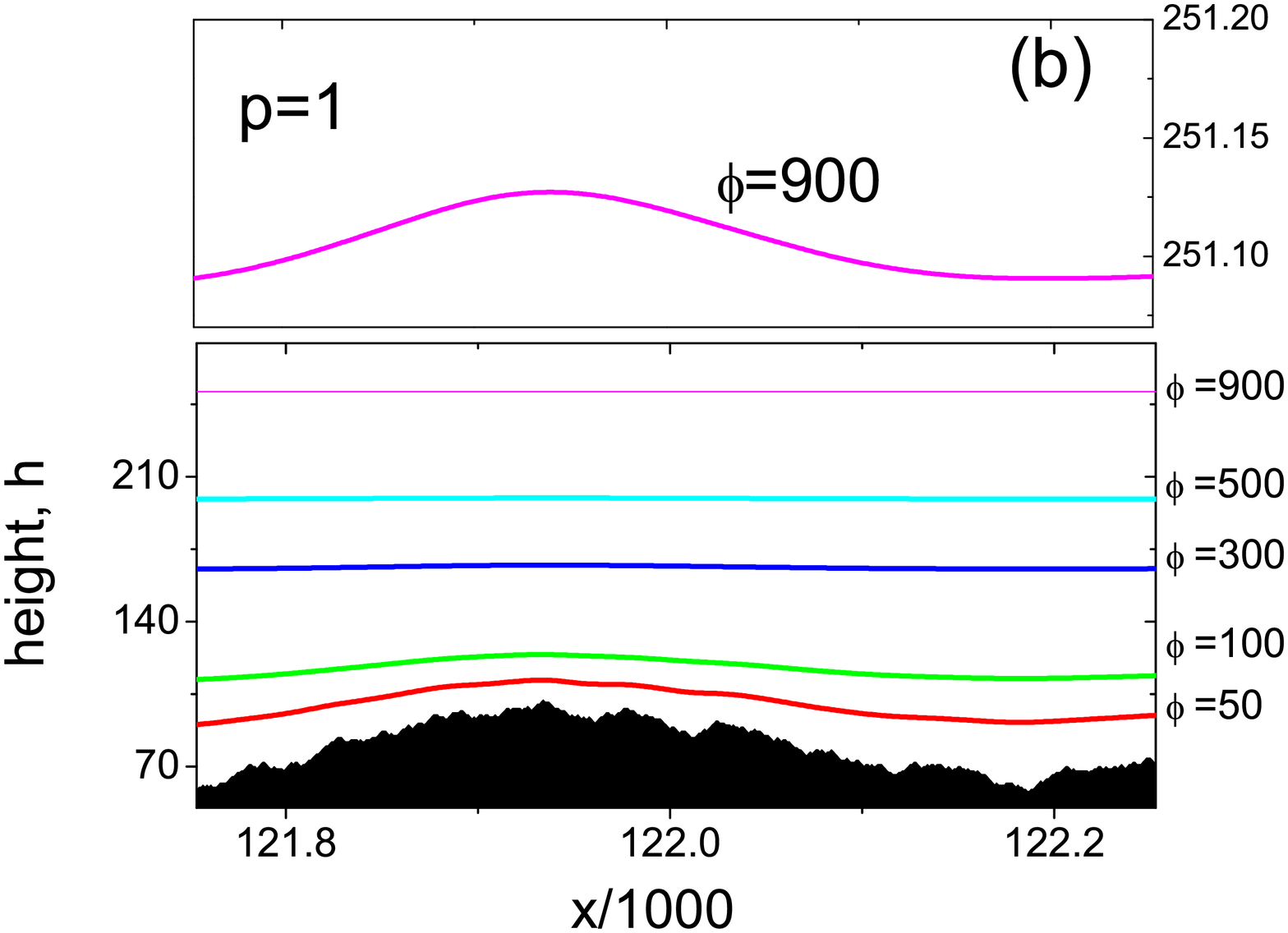}
\caption{Geometry of the problem for a substrate that was formed
using an irregular profile, which was defined by (a) p=0 and (b)
p=1. The equipotential lines $h_{p,\phi}(x)$ for $\phi = 50, 100,
300, 500$ and $900$ are shown. The top panel shows the equipotential
line $h_{p,900}(x)$ in a magnified scale.} \label{Figure1}
\end{center}
\end{figure}

Then, the domain is divided into a two-dimensional grid, and the
potential is iteratively calculated at each grid point for fixed
boundary conditions at the rough cathode, which is described by the
function $h_{p,0}(x)$ and the anode at $h_{p,A}$. Periodic boundary
conditions are imposed on the lateral borders. To determine the
coordinates of the equipotential lines, we define the following
difference,

\begin{equation}
D_{\phi} = \frac{\phi_{x,y+1} - \phi_{x,y}}{\Delta y} > 0,
\label{Eq3}
\end{equation}
where $x,y$ are integer numbers, and ${\Delta y}$=1 $\ll L$. For any
equipotential line $h_{p,\phi}(x)$ where
$\phi_{x,y}<\phi<\phi_{x,y+1}$, the corresponding coordinates are
computed by evaluating $h_{p,\phi}(x) = h(x) + dh_{\phi}$, with

\begin{equation}
dh_{\phi}= {D_{\phi}^{-1}}[\phi - \phi_{x,y}].
\label{Eq4}
\end{equation}
To illustrate the geometry of our problem, Figure \ref{Figure1}
shows the substrate (or cathode) and a family of equipotential lines
that were calculated using Eqs. (\ref{Eq2}), (\ref{Eq3}) and
(\ref{Eq4}) for $p=0$ (a) and $p=1$ (b).

Next, we characterize the height distributions of the corresponding
equipotential lines. In particular, the skewness, $S(\phi)=  L^{-1}\sum\limits_{x=1}^{L} \left(h_{p,\phi}(x) - \overline{h}_{p,\phi}\right)^{3}/W^{3}$ and kurtosis $K(\phi)= \left[L^{-1} \sum\limits_{x=1}^{L} \left(h_{p,\phi}(x) - \overline{h}_{p,\phi}\right)^{4}/W^{4}\right] - 3$ are used throughout this work.
Here, $W$ is a shorthand notation for $W_{p,0}(L,t)$, which is
defined in Eq. (\ref{Eq2a}).

\section{Results and Discussion}
\label{SECII}

\begin{figure}[!htb]
\begin{center}
\includegraphics [width=7.5cm] {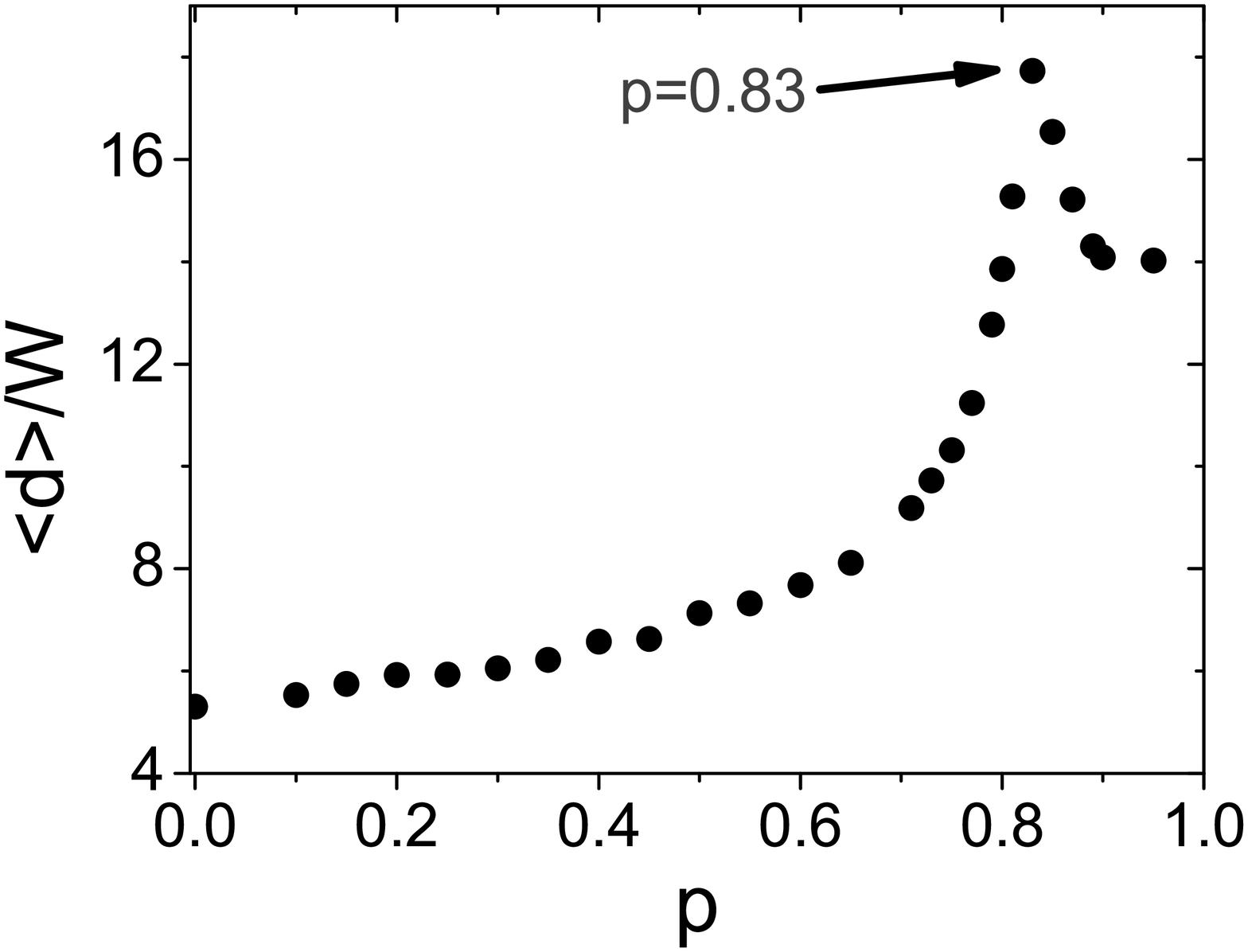}
\caption{Ratio $\frac{\langle d \rangle}{W}$ as function of
probability $p$. $\langle d \rangle$ is the average
distance between the anode and the irregular cathode (see Eq.\ref{EqE}) while
$W\equiv W_{p,0}(L,t)$ is the roughness of the irregular
cathode (see Eq.\ref{Eq2a}). The results are presented considering $\langle d \rangle$=198, and it is possible to observe that the global roughness
decreases at the interval $0<p<0.83$ and increases for $p>0.83$. The
global roughness is minimum at the point $p=0.83(\approx p^{*}$)}.
\label{larg}
\end{center}
\end{figure}

Before discussing the results for the Family and the competitive
models, we emphasize that in this last case, our results are based
on a single profile for each distinct value of $p$. This usage is
justified by the following facts: for each profile, our procedure
amounts to considering an entire set of equipotential lines, which
demands a considerably numerical effort; we have found that, in the
interval $p\lesssim 0.75$ and $p \gtrsim 0.9$, at time $T$, the
height distribution of the cathode and the resulting equipotential
lines are almost insensitive to $p$, which can be observed by
comparing the values of $S$ and $K$. Therefore, our procedure is
equivalent to replacing several realizations for one value of $p$
with a set of independent realizations at slightly different values
of this parameter. This method can also be justified when we move to
the close neighborhood of $p=p^*$. Here, $S$ and $K$ vary smoothly
even for the results that are produced using 1 profile realization.

The first illustrated aspect in Figure \ref{larg} refers to the
relative positions of the profile with respect to the anode, which can be expressed by
the ratio $\frac{\langle d \rangle}{W}$ ($\langle d \rangle = 198$) as a function of $p$.
In units of the roughness $W$, the curve in Figure \ref{larg} represents the height where
the anode is maintained above the average height of the
corresponding rough profile. It is possible to observe that the
smoother rough profile occurs at $p=0.83$, i.e., near the critical
point. Moreover, it is expected that in the infinite-size limit, the peak
becomes more pronounced.

\begin{figure}[!htb]
\begin{center}
\includegraphics [width=7.0cm] {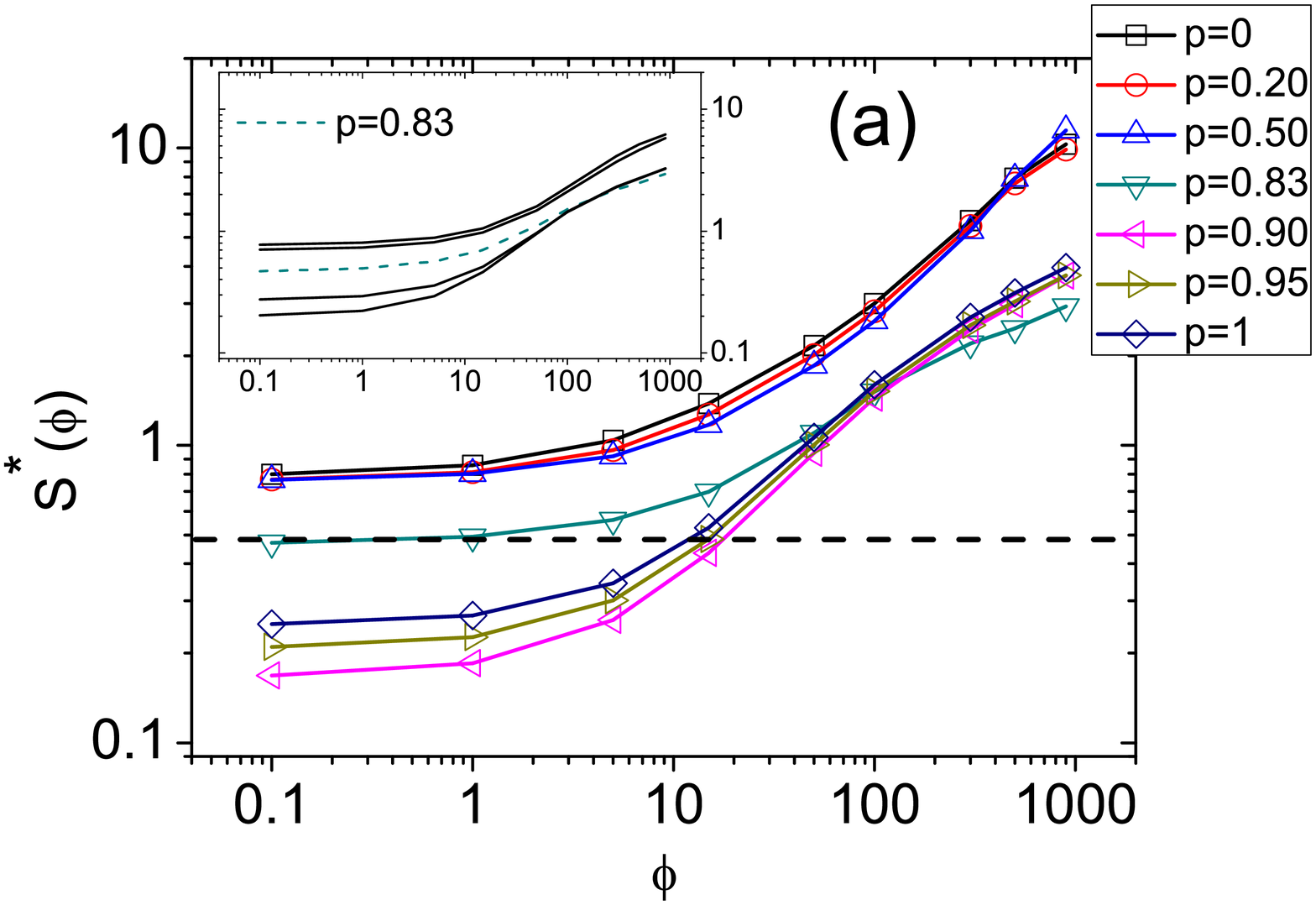}
\includegraphics [width=7.0cm] {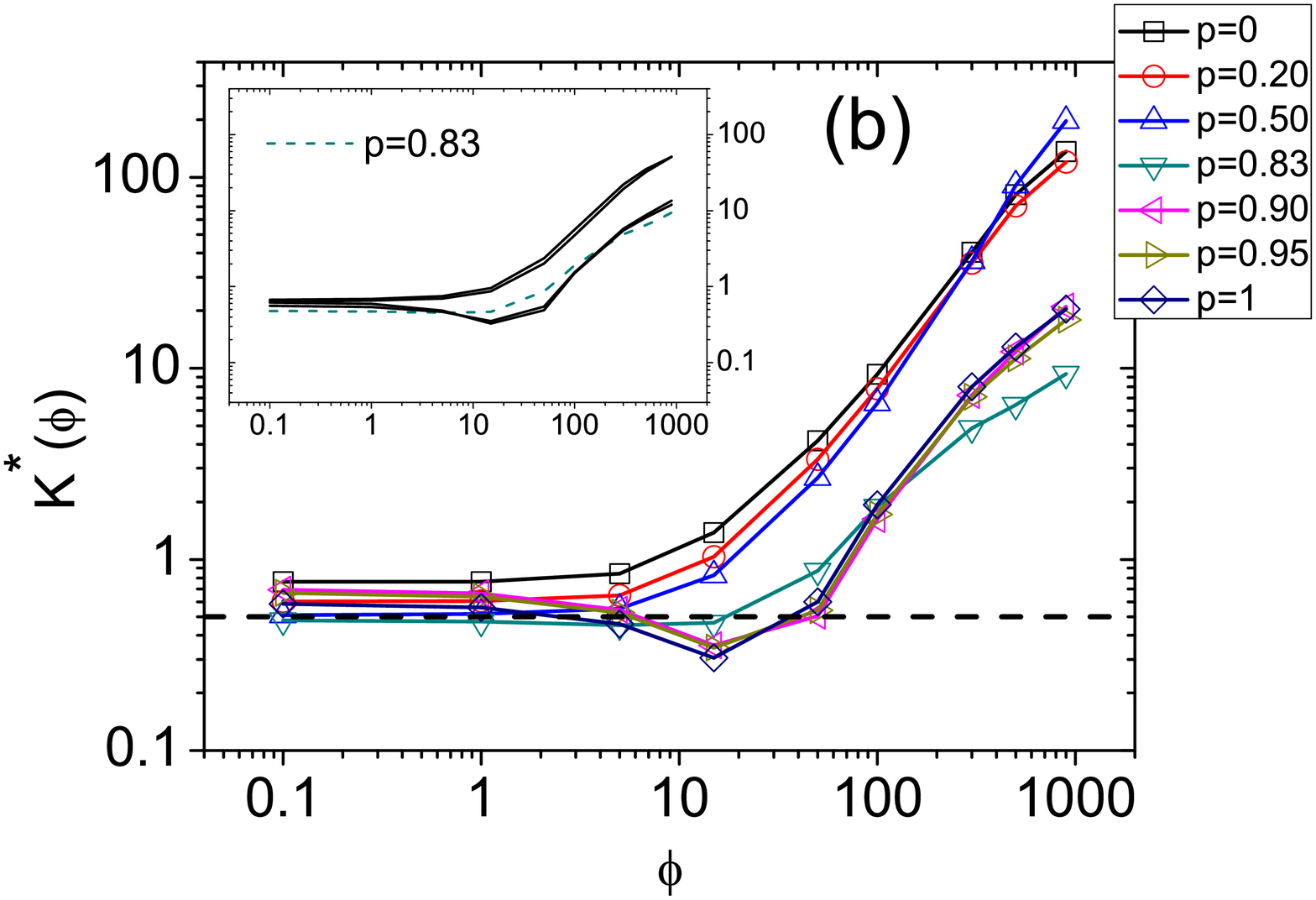}
\includegraphics [width=7.0cm] {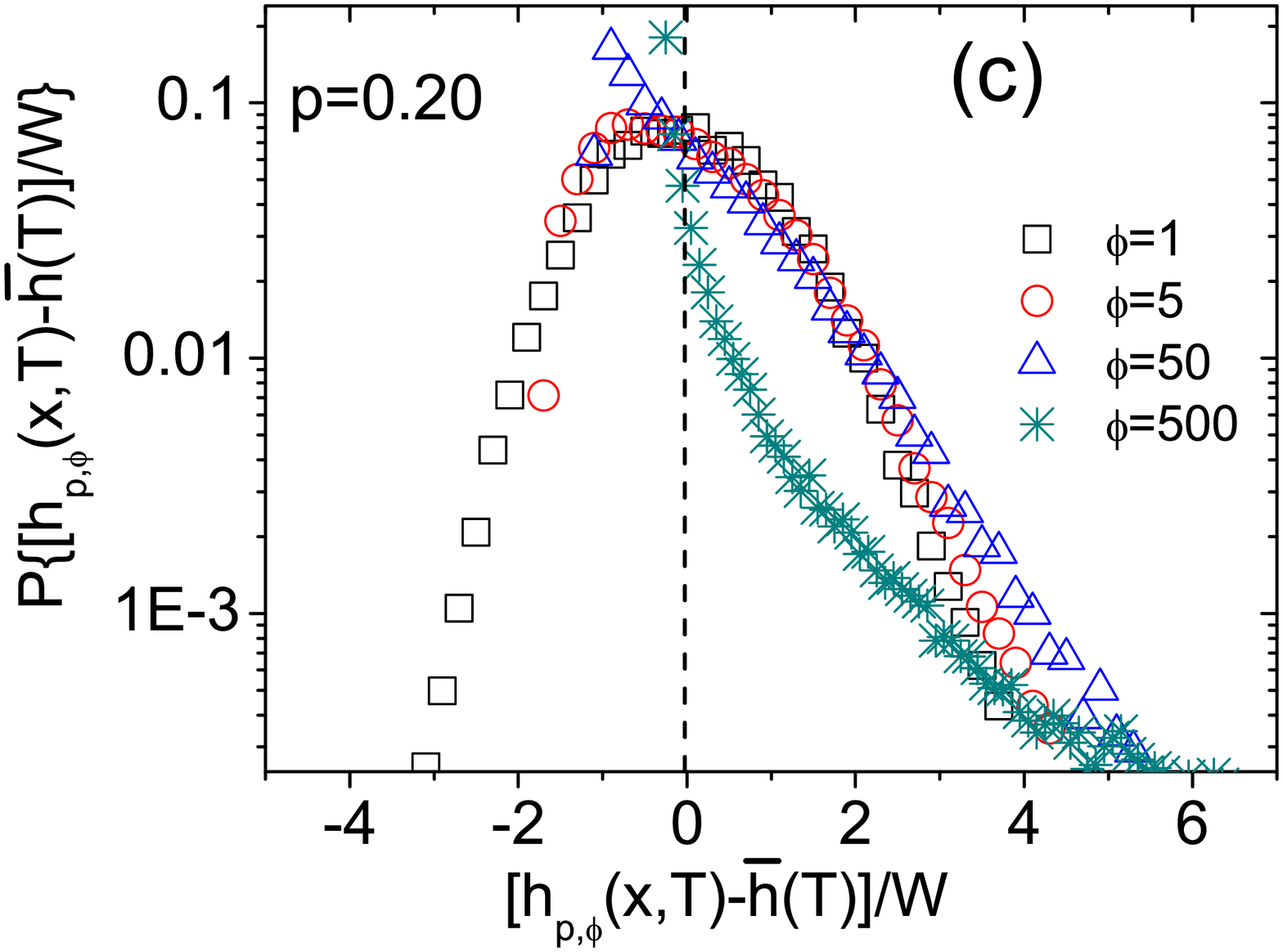}
\caption{Shifted skewness $S^*$ (a) and kurtosis $K^*$ (b) for a
family of equipotential lines $h_{p,\phi}(x)$ as a function of
$\phi$ for different values of $p$. Three distinct regimes can be
identified: for $\lambda>0$ $\Leftrightarrow$ $p=0, 0.20$ and
$0.50$, $\lambda\approx0$ $\Leftrightarrow$ $p=0.83$, and
$\lambda<0$ $\Leftrightarrow$ $p=0.90, 0.95$ and $p=1.0$. The inset
details the behavior in a region where the rough profile
morphologies are blurred because of finite time effects. The
horizontal dashed line defines $S^{*}(\phi)=0.5 \Rightarrow
S(\phi)=0$ and $K^{*}(\phi)=0.5 \Rightarrow K(\phi)=0$. (c) Scaled
height distributions for the equipotential lines considering $\phi =
1,5,50$ and $500$ for a rough conducting profile grown with
$p=0.20$. The vertical dashed line defines $\left[h_{p,\phi}(x,T)-
\overline{h}(T)\right]/W = 0$. } \label{skew}
\end{center}
\end{figure}
Next, we characterize the height distribution of the equipotential
lines based on the behavior of $S$ and $K$ as a function of $\phi$.
These quantities vary in a wide interval, so that it becomes
convenient to draw some of the graphs in the logarithmic scale.
Because both quantities may assume negative values, we shift $S$ and
$K$ by a fixed value $0.5$, which leads to the functions
$S^{*}(\phi) = S(\phi) + 0.5$ and $K^{*}(\phi) = K(\phi) + 0.5$
illustrated in Figure \ref{skew}. To better interpret the results,
we draw the horizontal lines $S^{*} = 0.5$ and $K^{*} = 0.5$, which
delimitate the regions of negative and positive values of $S$ and
$K$. The overall behavior of $S^{*}(\phi)$ is characterized by its
monotonic increase with respect to the value of $\phi$, as displayed
in panel (a). This result is consistent with the convergence of $S$
to a $\delta$ function, which is expected for a flat equipotential
line. Furthermore, for $p\lesssim0.75$, the results in Figure
\ref{skew} reveal that $S(\phi\lesssim10)\sim 0.29$ (a) and
$K(\phi\lesssim10)\sim 0.16$. Therefore, in the growth regime where
the fluctuations in the surface height distribution is described
using the GOE-TW distribution, the equipotential lines close to the
profile follow a similar height distribution. The values of
$S^{*}$ and $K^{*}$ do not significantly change until $\phi \approx
10$. However, for $\phi \gtrsim 10$, the derivative $dS/d\phi$
significantly grows, which indicates a non-uniform convergence to
the $\delta$ function. This aspect can be observed by the analysis
of Figure \ref{skew} (c), which shows the scaled height distribution
$P\left\{[h_{p,\phi}(x,T)- \overline{h}(T)]/W\right\}$ of the
equipotential lines for $\phi = 1,5,50$ and $500$. In fact, our
results show that the large deviations of $\overline{h}(T)\equiv\overline{h}_{p,\phi}(T)$
for each equipotential line determine the behavior of $S$. For $p =
0.83$, the behavior of $S(\phi)$ has similar but less pronounced
features, which indicates a weaker dependence of $dS/d\phi$ at this
value of $\lambda$.

\begin{figure}[!htb]
\begin{center}
\includegraphics [width=8.5cm] {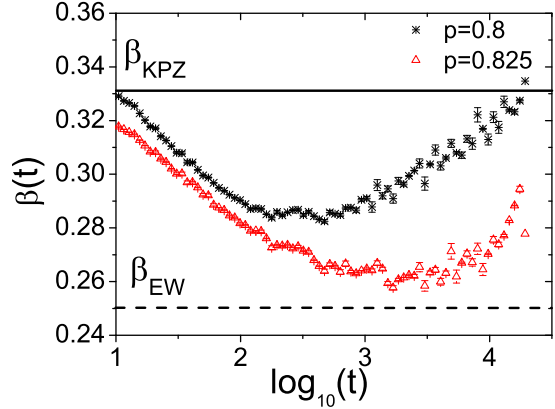}
\caption{Dependence of the local growth exponent $\beta(t) (\equiv d\ln{W}/d\ln{t})$ for
$p=0.8$ and $p=0.825$. The dashed and solid lines indicate the exact
EW and KPZ values.} \label{local}
\end{center}
\end{figure}

For $p>0.90$, $S(\phi)$ becomes negative for $h_{p,\phi}(x)$ near
the rough conducting profiles $h_{p,0}(x)$. In this region, the
absolute value of $S$ is also near that typical of a GOE TW
distribution. The important aspect is that $S(\phi)$ crosses the
null value at $\phi \approx 15$, which coincides with the point
where $K^{*}(\phi)$ exhibits a minimum value (see Figure \ref{skew}
(b)). This result indicates that, for conducting surfaces where
$\lambda<0$, there is an equipotential line near the surface that
presents a symmetric asymptotic behavior at the tails of the height
distributions. This result suggests that the height distribution of
the equipotential lines changes in a different manner (for positive
and negative fluctuations) with respect to that of the actual
conducting profile.

In the insets of Figures 3 (a) and 3 (b), we focus on the behavior
of $S^*(\phi)$ and $K^*(\phi)$ for p = $0.80$, $0.81$, $0.87$ and $0.89$,
where for $L$ and $t=T$, the system presents morphologic features in
the crossover EW-KPZ. Interestingly, the results indicate that
similar features do not appear for the values of $S^*(\phi)$ and
$K^*(\phi)$. We notice no appreciable differences for $t=T$, in the
values of $S^*(\phi)$ and $K^*(\phi)$ for $p$ $\in$ [0, 0.75] and
$p$ $\in$ (0.75, 0.83) or $p$ $\in$ (0.83, 0.9) and $p$ $\in$ [0.9,
1]. Our results suggest that the potential variation does not feel
the consequences of these finite time effects (KPZ-EW crossover),
following the typical values for models that obey the KPZ
statistics. The analysis of equipotential lines appears as a better
estimator to uncover the actual geometric features of the growing
system.

Let us proceed with a closer discussion of our results for $p \in
(0.75, 0.83)$ and $p \in (0.83, 0.9)$ at times where the EW features
are believed to exist. It is well known \cite{Caio,Oliveira} that in
this region, the duration of the transient phase EW-KPZ increases
quickly. More specifically, it is claimed that the time to
reach the asymptotic limit (i.e. the KPZ class) scales with
$\lambda^{-4}$ \cite{Oliveira2}. In competitive models, $\lambda$
grows at least linearly with the difference $p-p^{*}$
\cite{Silveira}, and the crossover time is expect to depend on $p$
with $(p-p^{*})^{-4}$. This result is corroborated by evaluating the
local growth exponent $\beta(t) \equiv d\ln{W}/d\ln{t}$ (or effective growth exponent)
illustrated in Figure \ref{local}. Let us first recall that, for any
growing profile, a single growth exponent $\beta$ can only be
defined if a strict linear dependence between $\log(W(L,t))$ and
$\log(t)$ is observed. Because this dependence is not verified for a
large interval of values of $t$ when $0.75 \lesssim p \lesssim 0.9$,
we analyze the behavior of $\beta(t)$, which is computed by
numerically evaluating the local slope of this curve. We consider
two conditions for which linear features should prevail in the
morphology of conducting rough profiles as $p=0.8$ and $p=0.825$.
Figure \ref{local} shows the time evolution of $\beta(t)$, which
indicates the presence of a minimum that hints the value of $t$
where the profile should be closer to one grown using a pure EW
dynamics (this minimum was estimated at the times $t=347$ and
$t=1652$ for $p=0.8$ and $p=0.825$, respectively).

\begin{figure}[!htb]
\begin{center}
\includegraphics [width=7.5cm] {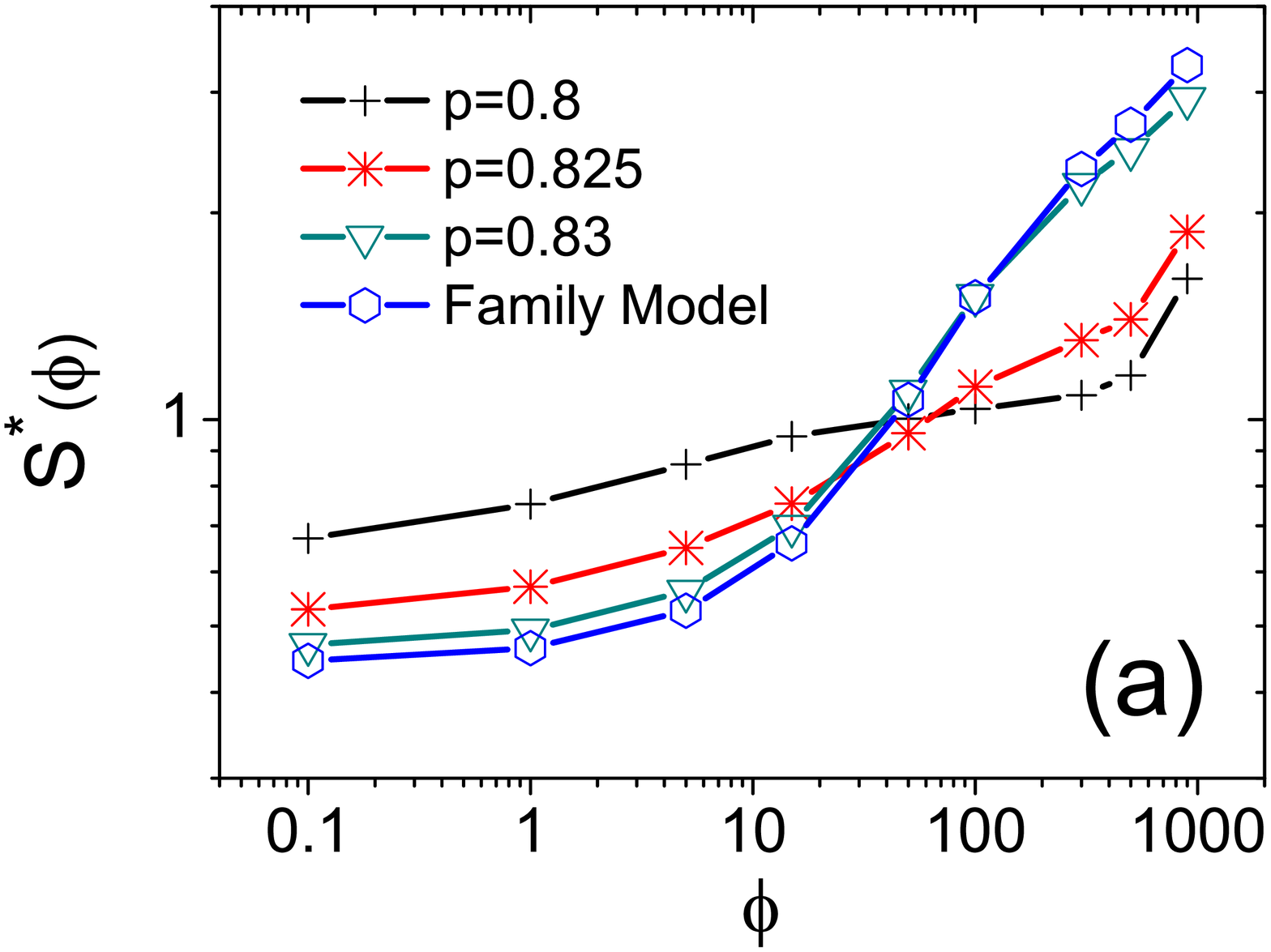}
\includegraphics [width=7.5cm] {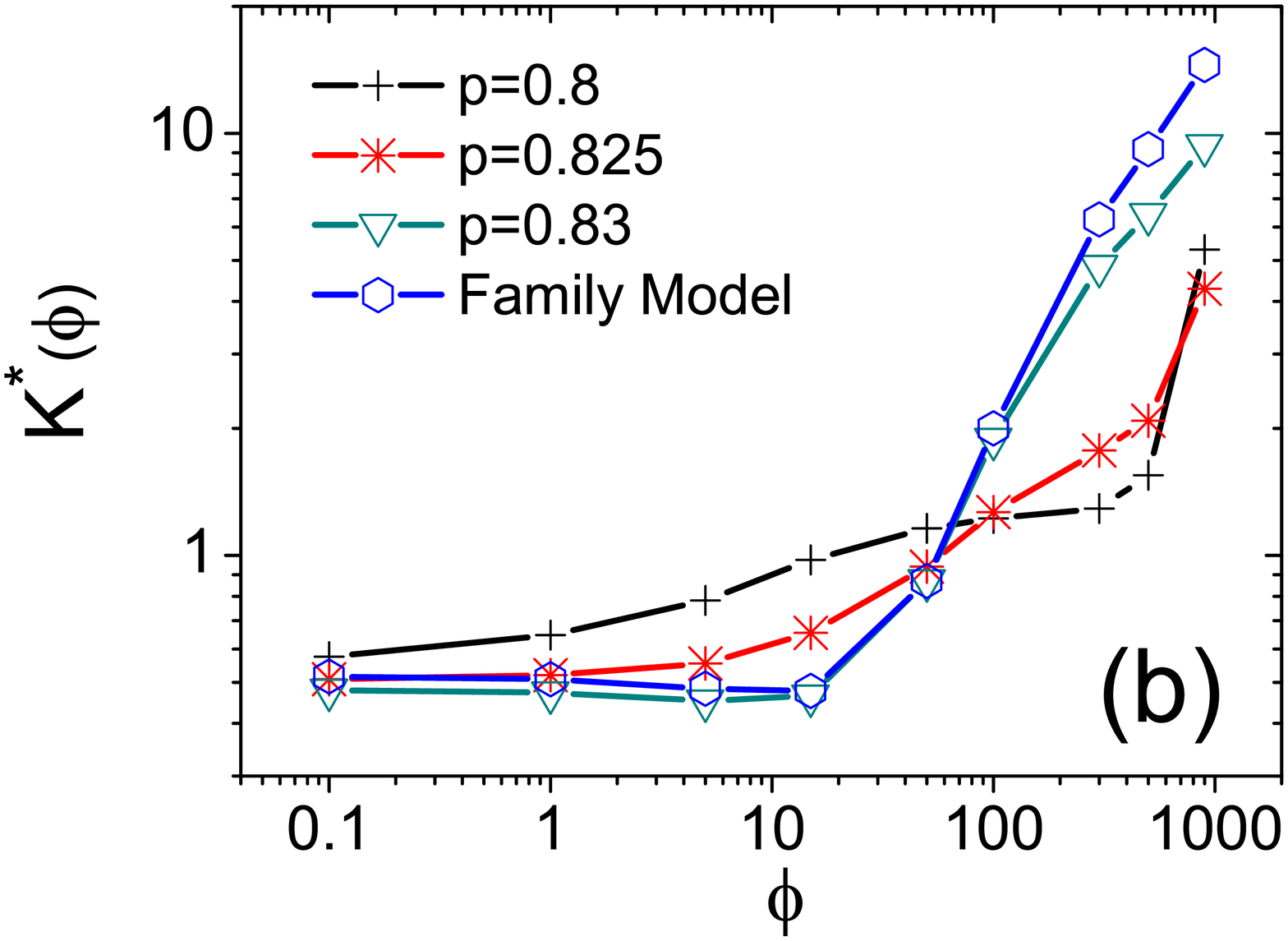}
\caption{Shifted (a) skewness $S^*$ and (b) kurtosis $K^*$ of
equipotential lines as functions of the electric potential for the
rough profiles that were grown according to the Family model and the
competitive model at $p=0.8$, $p=0.825$ and $p=0.83 \approx p^{*}$.
For $p=0.8$ and $p=0.825$, the rough cathode surfaces were
considered at times when the $\beta(t)$ is minimum (See Figure
\ref{local}). This minimum was estimated when $t=347$ and $t=1652$
for $p=0.8$ and $p=0.825$, respectively. It is possible to observe
the convergence to EW behavior in the  $p\rightarrow p^{*}$ limit.}
\label{skewfm1}
\end{center}
\end{figure}

To show that the analysis of equipotential lines helps elucidating
the fine differences between profiles, we also explicitly consider a
rough profile that is grown using the well-know linear Family model.
Then, we compare the resulting behavior of the corresponding
equipotential lines with those for the competitive model for
$p=0.83\approx p^*$ and $p\sim 0.83$ ($p=0.8$ and $p=0.825$ at times
$t=347$ and $t=1652$, respectively). In Figure \ref{skewfm1} (a) and
(b), we observe a similar behavior for $S^*$ and $K^*$,
respectively, as a function of $\phi$ for the Family model and as
$p=0.83\approx p^*$. This result is interesting because modeling the
condition $p=0.83$ provides vacancies in the volume, but these
features are not reflected in the behavior of the equipotential
lines, as we can observe by directly comparing with the results
obtained using the Family model. However, for $p\sim 0.83$, we do
not observe a similar behavior for the equipotential lines, which
now exhibit different morphologies from the previous ones. This
result indicates that $\beta(t)$ of the competitive model close to
those that characterize the EW dynamic (minimum of $\beta(t)$), is
not a reliable measure to confirm the electrical features of the
rough devices that are similar to that grown using the EW
dynamics. Moreover, we observe that for $p=0.8$, $S^*$ exhibits
low values for far equipotential lines, compared with the other models
($S^{*}\sim 1$ for $\phi=900$). Finally we can observe that a
notable convergence for the limit $p\rightarrow p^{*}$ and for
values of $t$ that correspond to the minimum of $\beta(t)$, to the
behavior of $S^*$ and $K^*$ of equipotential lines, which are
associated to EW features. At these values of $t$, the
misinterpretations of small differences in $\beta(t)$
compared to $\beta_{EW}$ can significantly
affect the design of electrical devices, where roughness is an
important tool.

At this point, let us remark that the height distribution of the
equipotential lines becomes a reliable criterion to analyze the
morphological differences that are produced by distinct rough
cathodes. Indeed, the previous results, which were only based on the
local roughness exponent (see reference \cite{Hugo}), have not
produced such clear-cut differences in the local roughness exponent
of the equipotential lines that are produced by rough cathodes
generated by a ballistic ($\lambda
>0$) and random deposition with surface relaxation (RDSR) ($\lambda
=0$). In the current work, such differences become clear when we
consider the height distributions of equipotential lines for all
three regimes, which are characterized by $\lambda$ ($>0,=0,<0$) and
$p$ ($<p^{*},=p^{*},>p^{*}$).

Further morphological aspects of the profile have been revealed by
our results. Let us define the set of peaks $\mathcal{P}_{p,\phi}$
in an equipotential line. A point $x \in \mathcal{P}_{p,\phi}$,
where its associated vertical coordinate is larger than those of its
first lateral neighbors, i.e.,
$h_{p,\phi}(x-1,T)<h_{p,\phi}(x,T)>h_{p,\phi}(x+1,T)$. In a similar
way, let us define $\mathcal{F}^+_{p,\phi}$, the set of peaks of
an equipotential line that are also characterized by positive
fluctuations. A point $x \in \mathcal{F}^+_{p,\phi}$ when $x \in
\mathcal{P}_{p,\phi}$ and $h_{p,\phi}(x,T)>
\overline{h}_{p,\phi}(T)$. Finally, let us define $f_{+}$ as the
fraction of the number of points $x\in \mathcal{P}_{p,\phi}$ that
also satisfy $x\in \mathcal{F}^+_{p,\phi}$. It is possible to show
that a relation $\overline{h^3}_{p,\phi}(T) \sim
(\overline{h}^{+}_{p,\phi}(T))^{3} f_{p}$ should be valid, where
$\overline{h}^{+}_{p,\phi}(T)$ is the average value of
$h_{p,\phi}(x,T)$ that was restricted to points $x\in
\mathcal{F}^+_{p,\phi}$.

\begin{figure}[!htb]
\begin{center}
\includegraphics [width=7.5cm] {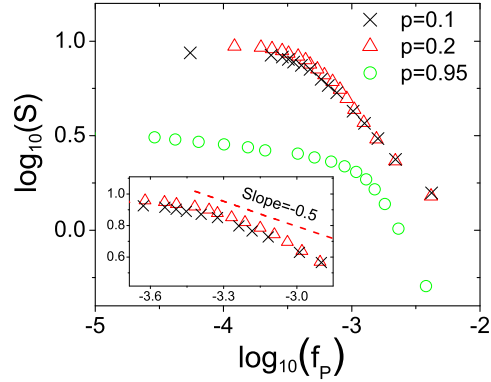}
\caption{Skewness as a function of the fraction of peaks for
equipotential lines that are calculated in the region bounded by a
rough profile and the flat line for $p= 0.10, 0.20$ and $0.95$. In
the inset, we show a magnified region where the scale relation $S
\sim f_{P}^{-1/2}$ holds.} \label{skew1}
\end{center}
\end{figure}

This regime identifies the regions for equipotential lines where the
small fluctuations quickly disappear and large fluctuations with
lower wavelengths prevail when the electric potential grows. In
fact, for equipotential lines that are far from the rough profile,
we find that $ \overline{h}^{+}_{p,\phi}(T) >>
\overline{h}_{p,\phi}(T) $ (For example, the main peaks in Figure
\ref{Figure1} (a) for $\phi=300$ and $500$). Therefore, the
corresponding global roughness can be written as

\begin{equation}
W^{3} \sim f_{+}^{-3/2} (\overline{h}^{+}_{p,\phi}(T))^{3} .
\label{Eq4m}
\end{equation}
Additionally, $S$ should depend on $f_+$ as:

\begin{equation}
S \sim f_{+}^{-1/2},
\label{Eq4n}
\end{equation}
so that $S>>1$ for equipotential lines that are far from the
cathode. This result is observed in Figure \ref{skew} (a), where
$S^{*}\sim 10$ as $\phi=900$ for $p<0.83$. It is worth noting that
such behavior is observed in the large roughness limit ($p<<1$) of
the conducting profile. In Figure \ref{skew1}, we show that these
theoretical predictions are confirmed by numerical calculations for
the rough conducting profile that were generated for $p=0.1$ and
$0.2$. It is clearly observed that in the regime where
$\overline{h}^{+}_{p,\phi}(T)
>>  \overline{h}_{p,\phi}(T) $, the exponent of $f_{P}$ tends to
$-1/2$. In this region, $dS/d\phi$ quickly grows for $300<\phi<500$
immediately before the inflexion point (see Figure \ref{skew} (a)).

If we only consider the cases where the equipotential lines have
$S>0$, the predictions by Eqs. (\ref{Eq4m}) and (\ref{Eq4n}) fail
for $p=0.95$, i.e., in the case of smooth surfaces, where the values
of the skewness of distant equipotential lines are $S^{*} \sim 4$ as
$p\rightarrow 1$ (For example, see Figure \ref{skew} (a) for
$p>0.83$). In that case, $dS/d\phi$ grows slowly before the
inflexion point mainly because of the variations of fluctuations,
which are associated to longer wavelengths (See Figure \ref{Figure1}
(b)). For $p=0.95$, the region before the inflection point
corresponds to $-3.0\lesssim \log_{10}(f_{P}) \lesssim-2.5$.
However, Figure \ref{skew1} shows that, in this region, the absolute
value of the exponent of $f_{+}$ is greater than $1/2$. This result
explains the rapid disappearance of roughness with lower wavelengths
(which do not corresponds to pronounced peaks). Then, we encounter
another regime with the absolute value of the exponent of $f_{+}$
less than $1/2$, which explains the limit $dS/d\phi \rightarrow 0$
in Figure \ref{skew} (a) for $p=0.95$.

\section{Summary and Conclusions}
\label{conclusion}

In this paper, we study the behavior of the height distributions of
the equipotential lines and focus the behavior of their skewness and
kurtosis. We used rough profiles that were grown following a
competitive model with KPZ components, which show opposite signals
of $\lambda$. Our results show a non-trivial behavior of $S$ and $K$
as a function of $\phi$. We conducted our investigation for profiles
with a lateral extension $L=10^6$ grown until a time
$t=2\times10^{4}$, which considerably reduces (but not eliminate)
the presence of finite time effects for any $p\neq$ 0.83.

For all KPZ profiles with $\lambda>0$, an almost invariant behavior
of $S$ and $K$ on $\phi$ was found. The same behavior occurs to
families for which $\lambda<0$, where $S$ and $K$ for the height
distribution of the equipotential lines have an invariant behavior
with respect to $\phi$, although they are different from the
previous families with $\lambda>0$. For families with $\lambda
\approx 0$, our results show a third distinct behavior for $S$ and
$K$. These results suggest that the electric potential variation can
identify the signal of $\lambda$, which provides the signal of the
skewness of the rough conducting profile. However, the height
distribution statistics of equipotential lines far from the
conducting profile were shown as being notably different.

The particular behavior of $S$ and $K$ as a function of $\phi$ when
$p\approx p^{*}$ motivated the comparison of these parameters with
the corresponding ones that were obtained for substrates that are
characterized by identical EW values of growth exponents. This
comparison was conducted considering the equipotential lines of
rough profiles: (i) which are defined by $p=0.83$, (ii) for the case of a
Family model, and (iii) at the time in the competitive model (for $p$=0.80
and $p$=0.825) where the effective growth exponent nearly $1/4$. The
results show similar behaviors for $p=0.83$ and the Family model but
clearly different results when the effective growth exponent is
nearly $1/4$. At these values of $t$, where thin-film technology arises,
the misinterpretations of small differences in $\beta(t)$
(if experimentally reliable) compared with
$\beta_{EW}$ can significantly affect the design of the electrical
devices.

Finally, our results can provide new insights for cases where the
rough morphologies are an important tool to construct devices where
electrical properties play a fundamental role. For example in the
design of devices for the flexible stacked resistive random access
memory (RRAM) applications the rough surface leads to a local
electric field enhancement which accelerates the oxygen ion
back-drift by the external positive bias (reset process)
\cite{Jeong}. Radio-frequency absorption and electric field
enhancements due to surface roughness are also subjects of
theoretical research \cite{Zhang}. However, there are situations where high
roughness is an important tool, such as when the deposited surface
is combined with metallic and nonmetallic materials. In that case, understanding
the evolution of the surface morphology and the electric potential
distribution along an irregular shape during electrodeposition
phenomena is greatly significant \cite{Gamburg}.

Moreover, in our viewpoint, the current study has called the
attention to interesting characteristics that help one more
systematically interpret of images from probe microscopies, such as
the Electrostatic Force Microscopy. In that
case, the tip must be maintained far enough from the rough surface
to avoid unwanted short-range interactions (which may result from
van der Waals forces) and close enough to not loose important
morphologic information about the real profile.


\section*{Acknowledgements}

This work has the financial support of CNPq and FAPESB (Brazilian
agencies). C. P. de Castro acknowledges specific support from
Funda\c{c}\~{a}o CAPES.




%

\end{document}